\documentclass[letterpaper, 10 pt, conference]{ieeeconf} 

\IEEEoverridecommandlockouts          
\usepackage{enumerate}
\usepackage{amsmath}
\usepackage{amssymb}
\usepackage{hyperref}
\usepackage{graphicx}
\usepackage{epstopdf} 
\usepackage{mathtools}
\usepackage{booktabs}
\usepackage[english]{babel}
\usepackage{xcolor, soul}
\usepackage{siunitx}    

\usepackage[firstpageonly=true]{draftwatermark}	

\newtheorem{remark}{Remark}
\newtheorem{assumption}{Assumption}

\newtheorem{proposition}{Proposition}

\newcommand\REV[1]{#1}
\newcommand\scalemath[2]{\scalebox{#1}{\mbox{\ensuremath{\displaystyle #2}}}}

\title{Towards lifelong learning of Recurrent Neural Networks for control design}

\author{Fabio Bonassi$^{1, *}$ \thanks{$^*$ Corresponding author}, Jing Xie$^{1}$, Marcello Farina$^{1}$, and Riccardo Scattolini$^{1}$
	\thanks{$^{1}$ The authors are with the Dipartimento di Elettronica, Informazione e Bioingegneria, Politecnico di Milano, Via Ponzio 34/5, 20133, Milano, Italy. E-mail: {\tt\small name.surname@polimi.it}}}

\begin{document}

\maketitle
\thispagestyle{empty}
\pagestyle{empty}

\begin{abstract}
This paper proposes a method for lifelong learning of Recurrent Neural Networks, such as NNARX, ESN, LSTM, and GRU, to be used as plant models in control system synthesis.
The problem is significant because in many practical applications it is required to adapt the model when new information is available and/or the system undergoes changes, without the need to store an increasing amount of data as time proceeds.
Indeed, in this context, many problems arise, such as the well known Catastrophic Forgetting and Capacity Saturation ones.
We propose an adaptation algorithm inspired by Moving Horizon Estimators, deriving conditions for its convergence.
The described method is applied to a simulated chemical plant, already adopted as a challenging benchmark in the existing literature.
The main results achieved are discussed.
\end{abstract}

 \DraftwatermarkOptions{%
 angle=0,
 hpos=0.5\paperwidth,
 vpos=0.96\paperheight,
 fontsize=0.012\paperwidth,
 color={[gray]{0.2}},
 text={
   \parbox{0.99\textwidth}{\copyright \, 2022 EUCA. This article appears in the Proceedings of the 2022 European Control Conference (ECC'22), July 12-15, 2022, London, pp. 2018-2023. \\ DOI: \href{https://doi.org/10.23919/ECC55457.2022.9838393}{10.23919/ECC55457.2022.9838393}.}},
 }

\section{Introduction}
Nowadays, one of the most up-to-date topics in Machine Learning (ML) research is the so-called \emph{lifelong learning}, see for instance the recent reviews  \cite{cossu2021continual, losing2018incremental, parisi2019continual, sodhani2020toward}.
The formulation of the problem, also called \emph{incremental learning} \cite{wu2019large}, \cite{van2019three}, \emph{never-ending learning} \cite{mitchell2018never},  or \emph{continual learning} \cite{cossu2021continual}, is quite intuitive.
Assume that a Machine Learning method is used to retrieve a model of a dynamical system, by means of suitable training and validation procedures.
Then, provided that new data are available, how and when should this model be updated, i.e. re-tuned?

The lifelong learning problem is particularly relevant when such a model is used to synthesize a model-based control law.
Indeed, even if the adopted control strategy guarantees closed-loop stability and robustness properties in the nominal case, should the plant exhibit time-varying characteristics, adaptation would typically be required to preserve the stability characteristics as well as to maintain performance throughout the plant life-span.
In this framework, two main cases of adaptation can be considered. \smallskip

\noindent \emph{Case 1:}
Over time, new data are collected, either under the same operating conditions as the data previously used for model training and testing, or at least under conditions of plant parameters slowly varying over time. These new data can be thus used to improve the model accuracy. \smallskip

\noindent \emph{Case 2:} the plant moves to new and unexplored operating conditions. That is, the data generation system significantly changes over time, and the existing model is unable to represent the new system behavior.  \smallskip

In order to face these cases, various life-learning ML techniques have been proposed to guarantee properly tuned and frequently updated models, without the need to store and process an increasingly larger amount of data.
Indeed, to deal with \emph{Case 1}, the obvious idea is to re-tune the model parameters based on the new available data, while for \emph{Case 2} it can be more advisable to enlarge the model structure, so as to improve the model capability to represent different operating conditions, which generally implies the necessity to re-train the model from scratch.

Both these approaches have, however, their own drawbacks.
Namely, concerning \emph{Case 1}, re-tuning the model when new data are available can lead to the so-called \emph{Catastrophic Forgetting} problem, which happens when the knowledge previously embedded in the model is discarded by the re-tuning procedure itself, with consequent reduced description capabilities, see  \cite{sodhani2020toward}.

As for \emph{Case 2}, selecting new model structures generally leads to a model which is unable of retaining the previous representation capabilities, let alone of well describing the new operating conditions.
This phenomenon is usually named \emph{Capacity Saturation}, see again \cite{sodhani2020toward}, and is especially found in many machine-learning models, for which a change in structure implies the need to repeat the entire training procedure.

Among the established models for dynamical systems control design, Neural Networks (NNs), and in particular Recurrent Neural Networks (RNNs). 
They owe their wide popularity in the control community to their ability to resemble dynamic models that can be proficiently used to design model-based control strategies, such as Model Predictive Control (MPC).
In this context, a popular family of RNN is the one of Neural Nonlinear AutoRegressive eXogenous (NNARX) networks, see \cite{levin1993control, levin1995identification, sastry1994memory, ali2015artificial, himmelblau2008accounts}, and \cite{piche2000nonlinear, hosen2011control, nagy2007model} for the joint use of NNARX models and MPC.
One of the main advantages of NNARXs is that their state correspond to past inputs and outputs, so that observers are not required to implement a state-feedback control law.
A second family of RNNs, characterized by a simple structure particularly easy to tune, is the one of Echo State Networks (ESNs), originally introduced in \cite{jaeger2001echo}, and subsequently used for control design in  \cite{pan2011model, ploger2003echo}.
However, in view of their excellent performance in a wide class of problems, the most popular RNN are the Long Short Term Memory (LSTM) networks \cite{hochreiter1997long, wu2019machine}.
Finally, Gated Recurrent Units (GRU) \cite{mohajerin2019multistep, rehmer2019using}, although characterized by a simpler structure than LSTMs, have been shown to achieve satisfactory performances in a number of applications \cite{bianchi2017recurrent}, and they are hence considered as a valuable, yet simpler, alternative to LSTMs.

Theoretical foundations for the use of NNARX, ESN, LSTM, and GRU for system identification and control have recently been laid in terms of Input-to-State Stability (ISS) \cite{sontag1995characterizations, jiang2001input} and Incremental Input-to-State-Stability ($\delta$ISS) \cite{angeli2009further, bayer2013discrete}, see \cite{bonassi2021nnarx, armenio2019model, bonassi2020lstm, bonassi2021stability, terzi2021learning}.
These stability notions have then been leveraged when using these networks for MPC design, see \cite{terzi2021learning, bonassi2021nonlinear}. \smallskip

In this context, the aim of this paper is to draw a research line for a possible solution to the lifelong learning for the previously described RNN architectures.
Specifically, \emph{Case 1} is considered, i.e. a model of the plant with constant or slowly-varying parameters.
To this end, we propose an approach based on the Moving Horizon Estimation (MHE), an optimization-based strategy which has been widely investigated and used by the control community, see \cite{rao2001constrained, RawlingsBook}.
In particular, the proposed method relies on the algorithm described in \cite{alessandri2008moving} -- suitably extended to cope with the specific problem at hand -- to tune the weights of the RNN model so as to improve its modeling performances.

This strategy is quite different from the solutions currently adopted by the machine learning community, which consist of ad-hoc NN structures and learning algorithms, see e.g. \cite{cossu2020continual, guo2020continual, mercangoz2020autonomous}.
Such strategies may indeed be well-suited to \emph{Case 2}, where new operating regions are progressively explored by the system, at the price of significantly complicating the control design stage. \smallskip

The paper is organized as follows.
In Section \ref{sec:problem} the problem is stated, and the  proposed algorithm for the periodic tuning and adaptation of the RNN is described.
Section \ref{sec:simulation} is devoted to present the application of the proposed method to a nontrivial simulation example, consisting of a nonlinear plant made by two chemical reactors and one separator. The example, taken from \cite{stewart2010cooperative}, is characterized by a strongly nonlinear behavior, and it is believed to represent a challenging benchmark for the class of problems here considered.
Finally, Section \ref{sec:conclusion} closes the paper with some conclusions and hints for future works.

\section{Problem formulation and adaptation algorithm} \label{sec:problem}
Consider a RNN model, belonging to one of the families NNARX, ESN, LSTM, or GRU, which can be described as a dynamical system in the following generic form \cite{terzi2021learning}:
\begin{equation} \label{eq:model}
	\begin{dcases}
		x_{k+1} =  f(x_{k},u_{k}; \Theta) \\
  		y_{k} = g(x_{k},u_{k}; \Theta)
	\end{dcases},
\end{equation}
where $k$ is the discrete time index, $\Theta$ is the vector of network's parameters, called weights, and $x$, $u$, $y$ are the state, input, and output vectors, respectively.
The input $u$ is assumed to lie in a compact set $\mathcal{U}$ containing the origin.
In addition, it is assumed for simplicity that the state $x$ is available; note that this is always true for NNARXs.

We herein assume that the real plant to be estimated is fully described by \eqref{eq:model} with $\Theta=\Theta^o$, where of course $\Theta^o$  is unknown \REV{and unique}.
In other words, we assume that \REV{no measurement noise affects the system, and that} the structure of the NN model coincides with that of the plant,  though its parameters must be properly tuned.
To this end, the algorithm described in the following, based on MHE, is proposed.

\subsection {RNN parameters' update algorithm for lifelong learning}
The proposed algorithm is intended to be run periodically every $N$ steps, where $N$ is a positive integer corresponding to the length of the time-window throughout which the data is collected from the system.
Thus, at $k = tN$, with $t \in \{1, 2, ..., \}$, an optimization problem is formulated, which seeks the parameters update that best explains the collected data.
In this setting, the model weights constitute an optimization variable, denoted by $\Theta_{k}$.
The underlying MHE optimization problem at $k = tN$ can be stated as

\begin{equation} \label{eq:mhe_formulation}
\begin{aligned}	
  \Theta_{k}^* = \arg&\min_{\Theta_{k}} \,\,  \Big\{ J_{k}(\Theta_{k}) = \sum_{i=0}^{N}\|y_{k-i}-\hat{y}_{k-i}\|^2 + \\
    & \qquad\qquad\qquad\qquad + \mu \|\Theta_{k}-\Theta^{*}_{k-N}\|^2 \Big\} \\
    \text{s.t.} & \quad \hat{x}_{k-N} = x_{k-N} \\
   & \quad \hat{x}_{k-i} = f(\hat{x}_{k-i-1}, u_{k-i-1}; \Theta_{k}) \\
  & \quad \hat{y}_{k-i} = g(\hat{x}_{k-i-1}, u_{k-i-1}; \Theta_{k}) \\
  & \quad  \forall i \in \{ 0, ..., N \}
  \end{aligned}
  \end{equation}
  where $\Theta^{*}_{k-N}$ denotes the optimal solution to \eqref{eq:mhe_formulation} at time $k-N$.
  Note that the cost function $J_k$ penalizes the mismatch between the past measured output $y_{k-i}$, with $i \in \{ 0, ..., N \}$, and that of the NN model, i.e. $\hat{y}_{k-i}$, obtained initializing \eqref{eq:model} in $x_{k-N}$ and simulating it with the same input applied to the plant.
  The second term of $J_k$ discourages significant deviations from the previously computed optimal solution $\Theta^{*}_{k-N}$.
  Hence, the coefficient $\mu$ defines the trade-off between the need to improve the model's performances ($\mu$ ``small''), and the necessity not to forget the information previously acquired, i.e. to avoid the catastrophic forgetting. 
  \REV{In this sense, $\Theta^{*}_{k-N}$ can be regarded as a prior for the optimization procedure at time $tN$.}
  \REV{The optimal solution $\Theta^{*}_{k}$ represents the updated set of weight of the model \eqref{eq:model}. At time $k=(t+1)N$ the optimization procedure is then repeated.}
  \smallskip


\subsection {Convergence properties}

Under the stated assumptions, the properties of the proposed MHE algorithm can now be proved along the lines of \cite{alessandri2008moving}.
To this end, let us define the function $\gamma_{-i}$ as the one that relates, for all $i \in \{ 0, ..., N \}$, the output $y_{k-i}$ to the initial state $x_{k-N}$ and to the sequence of previous inputs, denoted by $\bold{u}_{k-i} = \{ u_{k-N}, ..., u_{k-i} \}$.
These functions can hence be defined as the concatenation of the system's equations  \eqref{eq:model}
\begin{align}
	y_{k-i} &= \gamma_{-i}(x_{k-N}, \bold{u}_{k-i}; \Theta^o),\\
	\hat{y}_{k-i} &= \gamma_{-i}(x_{k-N}, \bold{u}_{k-i}; \Theta_k).
\end{align}
It is hence possible to define the vector of output reconstruction errors as
\begin{equation}
\scalemath{0.775}{
	\Gamma_{k}(\Theta^0,\Theta_{k}) = \begin{bmatrix}
		\gamma_{-N}(x_{k-N}, \bold{u}_{k-N}; \Theta^o) - \gamma_{-N}(x_{k-N}, \bold{u}_{k-N}; \Theta_k ) \\
		\vdots \\
		\gamma_{-1}(x_{k-N}, \bold{u}_{k-1}; \Theta^o) - \gamma_{-1}(x_{k-N}, \bold{u}_{k-1}; \Theta_k ) \\
		\gamma_{0}(x_{k-N}, \bold{u}_{k}; \Theta^o) - \gamma_{0}(x_{k-N}, \bold{u}_{k}; \Theta_k )
	\end{bmatrix}}.       
\end{equation}

%

We introduce the following assumption:
\begin{assumption}
There exist a \emph{K}\emph{-function} $\varphi$ such that
\begin{equation} \label{eq:existance_phi}
\|\Gamma_{k}(\Theta^0, \Theta_{k})\|^2\geq \varphi(\|\Theta^o-\Theta_{k}\|^2)
\end{equation}
and	a positive scalar $\delta$ such that, for any $\Theta^0$ and $\Theta_k$,
\begin{equation} \label{eq:definition_phi}
\frac{\varphi(\|\Theta^o-\Theta_{k}\|^2)}{\|\Theta^o-\Theta_{k}\|^2}\geq\delta.
\end{equation}
\end{assumption}

Letting
$$\varepsilon_{k}=\|\Theta^o-\Theta^*_{k}\|^2,$$
the main result can be stated. \smallskip

\begin{proposition} \label{prop:convergence}
If
\begin{equation}\label{eq:prop:condition}
\frac{2\mu}{\frac{1}{2} \mu+\delta}<1
\end{equation}
then $\varepsilon_{k}\rightarrow 0$ as $k\rightarrow \infty$.
\end{proposition}
\bigskip
\begin{proof}
Along the lines of \cite{alessandri2008moving}, let us first compute an upper bound of the cost function $J_k$ defined in \eqref{eq:mhe_formulation}, evaluated at the optimal value of the parameters $\Theta^*_{k}$.
To this end, consider $J_k(\Theta^o)$, and note that in this case $\|y_{k-i}-\hat{y}_{k-i}\|^2=0$, for all $i=0,...,N$,  since the model matches the plant.
This implies that
\begin{equation}
	J_{k}(\Theta^o)= \mu \|\Theta^o-\Theta^{*}_{k-N}\|^2 = \mu \, \varepsilon_{k-N}.
\end{equation}
Then, in view of the optimality of $\Theta^*_{k}$, one has that
\begin{equation}\label{eq:proof:opt}
	J_{k}(\Theta^*_{k}) \leq J_{k}(\Theta^o) = \mu \varepsilon_{k-N}.
\end{equation}
At this stage, let us recall that, by standard norm arguments, for any vector $z_a, z_b, \bar{z}$ it holds that
\begin{equation} \label{eq:proof:norm_properties_int}
\scalemath{0.85}{
\begin{aligned}
	\| z_a - z_b \|^2 \leq& \| z_a - \bar{z} \|^2 + \| z_b - \bar{z} \|^2 + 2 \big\| (z_a - \bar{z}) ( z_b - \bar{z}) \big\| \\
	\leq& \| z_a - \bar{z} \|^2 + \| z_b - \bar{z} \|^2 + \| z_a - \bar{z} \|^2 + \| z_b - \bar{z} \|^2 \\
	\leq& 2 \| z_a - \bar{z} \|^2 + 2 \| z_b - \bar{z} \|^2 ,\end{aligned}}
\end{equation}
where Young's inequality has been applied with $p=q=1$.
Thus, from \eqref{eq:proof:norm_properties_int} it follows that
\begin{equation} \label{eq:proof:norm_properties}
	\| z_a - \bar{z} \|^2 \geq \frac{1}{2} \| z_a - z_b \|^2 - \| \bar{z} - z_b\|^2.
\end{equation}

This inequality can be used to retrieve a bound of $J_{k}(\Theta^*_{k})$.
Taking $\Theta^*_{k}$, $\Theta^*_{k-N}$, and $\Theta^0$ as $z_a$, $\bar{z}$, and $z_b$, respectively, the second term of $J_{k}(\Theta^*_{k})$ can be bounded as:
\begin{equation}
	\begin{aligned}
  \mu \|\Theta^*_{k}-\Theta^*_{k-N}\|^2 \geq& \frac{\mu}{2} \|\Theta^*_{k} - \Theta^o \|^2 -\mu\|\Theta^*_{k-N}-\Theta^o\|^2 \\
    &= \frac{1}{2} \, \mu \, \varepsilon_{k} - \mu \, \varepsilon_{k-N}
\end{aligned}
\end{equation}

The first term of $J_{k}(\Theta^*_{k})$, i.e.  $\sum_{i=0}^{N} \|y_{k-i}-\hat{y}_{k-i}\|^2$, by definition can be expressed as
$ \| \Gamma_{k}(\Theta^0, \Theta_{k}^*)\|^2$.
Hence, in light of \eqref{eq:definition_phi}, $J_{k}(\Theta^*_{k})$ satisfies
 \begin{equation}\label{eq:proof:J_bound}
J_{k}(\Theta^*_{k}) \geq \frac{1}{2} \, \mu \, \varepsilon_{k} - \mu \,\varepsilon_{k-N} + \delta \, \varepsilon_{k}
\end{equation}

By combining \eqref{eq:proof:opt} and \eqref{eq:proof:J_bound}, we finally have that the dynamics of the the error $\varepsilon$ is governed by
\begin{equation}\label{eq:proof:err_dynamics}
	\varepsilon_{k} \leq \frac{2\mu}{\frac{1}{2}\mu+\delta} \varepsilon_{k-N}.
\end{equation}
Thus, if \eqref{eq:prop:condition} holds, the error converges.
\end{proof}

It is worth noticing that condition \eqref{eq:prop:condition} can always be satisfied provided that the tuning parameter $\mu$ is chosen sufficiently small, i.e. $\mu < \frac {2}{3}\delta$.
However, too small values of $\mu$ can lead the value of $\Theta^*$ to depend, to a large extent, on the most recent data (and possibly on noise), and ultimately leading to Catastrophic Forgetting, so that a suitable trade-off must be suitably selected. \bigskip

\begin{remark}
In the formulation of the adaptive algorithm described above, two fundamental assumptions have been implicitly introduced.
The first one is that there exist a \REV{unique} unknown parametrization $\Theta^{o}$ for which the RNN model perfectly describes the plant's dynamics.
Secondly, it has been assumed that the plants' parameters are constant, i.e. $\Theta^{o}$ is fixed.
However, the most natural framework to use the described method is the one where there is a plant/model mismatch, and the plant parameters are (very) slowly varying.
While for this setting the proposed theoretical results are not guaranteed to hold, the tests performed on the numerical example discussed in Section \ref{sec:simulation} show that these assumptions can in practice be relaxed.
\end{remark}

\section {Simulation example} \label{sec:simulation}
\begin{figure}[t]
\centering
\includegraphics[width=0.9\linewidth, clip, trim=0cm 5.25cm 9cm 0cm]{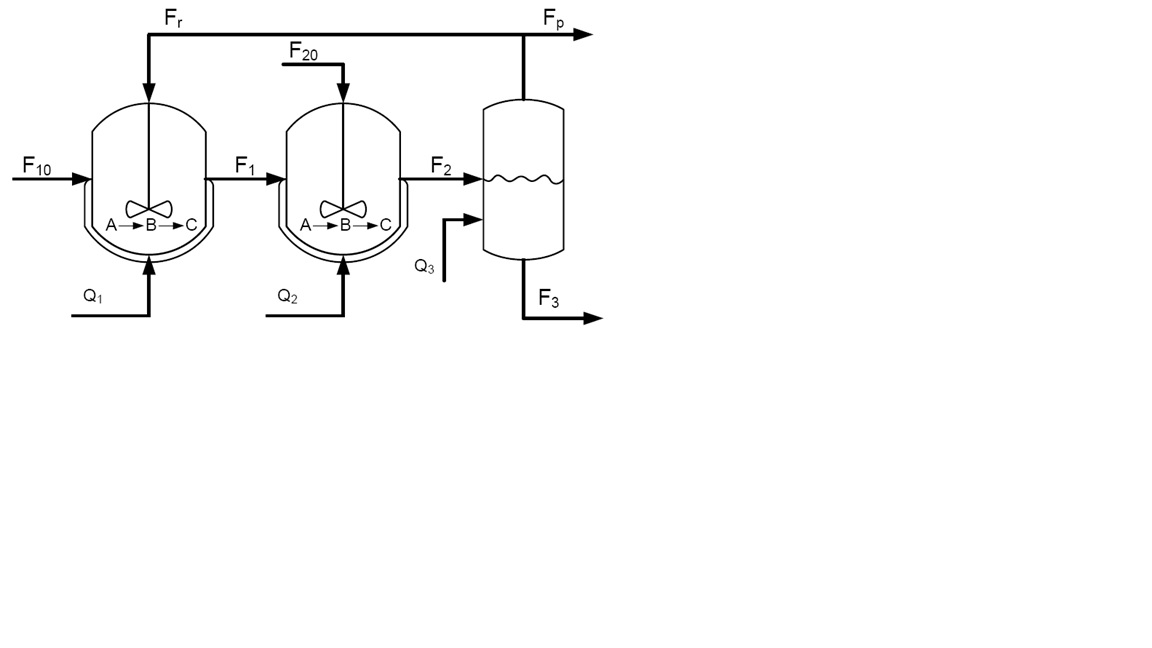}
\centering
\caption{Schematic of the considered chemical benchmark plant.}
\label{fig:reactors}
\end{figure}

\bigskip
To test the proposed strategy on a realistic problem, we applied the lifelong learning approach discussed in Section \ref{sec:problem} to an LSTM network describing the dynamics of the chemical benchmark system depicted in Figure \ref{fig:reactors}, which has been previously considered in \cite{stewart2010cooperative} and \cite{kumar2021industrial} as a challenging problem for control design and estimation.

\subsection{Chemical benchmark system description}
The chemical plant is composed of two reactors and one separator. 
The reactant liquid A feeds the two reactors, where it is converted into the products B and C. 
There is also a recirculation flow from the separator to the first reactor that makes the system strongly coupled.
The physical model of the plant is composed of twelve nonlinear continuous-time differential state equations, describing mass and energy balances.
These equations determine, for each reactor ($i=1$ and $i=2$) and for the separator ($i = 3$), the dynamics of the liquid's level $H_i$, its temperature $T_i$, and the concentrations of chemical products A and B, denoted by $x_{Ai}$ and $x_{Bi}$, respectively.
For simplicity, only the dynamics of the second reactor ($i=2$) are herein considered, by assuming that its inputs (partially coinciding with the outputs of reactor $1$) are independent terms.
The nonlinear model of  reactor $2$ is
\begin{equation} \label{eq:benchmark}
	\scalemath{0.875}{
		\begin{aligned}
			\dot{H}_{2\phantom{1}}=&\frac{1}{\rho A_2}(F_{20}+F_1-F_2)\\
			\dot{x}_{A2}=&\frac{1}{\rho A_2H_2}(F_{20}x_{A0}+F_1x_{A1}-F_2x_{A2})-k_{A2}x_{A2},\\
			\dot{x}_{B2}=&\frac{1}{\rho A_2H_2}(F_1x_{B1}-F_2x_{B2})+k_{A2}x_{A2}-k_{B2}x_{B2}, \\
			\dot{T}_{2\phantom{1}}=&\frac{1}{\rho A_2H_2}(F_{20}T_0+F_1T_1-F_2T_2)\\
			&-\frac{1}{C_p}(K_{A2}x_{A2}\Delta H_A+k_{B2}x_{B2}\Delta H_B)+\frac{Q_2}{\rho A_2C_pH_2},
	\end{aligned}}
\end{equation}
where
\begin{equation*}
\scalemath{0.9}{
\begin{aligned}
	F_i &=k_{vi}H_i \quad i \in \{ 1,2 \}, \\
	k_{A2} &= k_A\exp\left(-\frac{E_A}{RT_2}\right), \quad	k_{B2} = k_B\exp\left(-\frac{E_B}{RT_2}\right).
\end{aligned}}
\end{equation*}
For the considered system, the output vector $y \in \mathbb{R}^{4}$ coincides with the state vector $x \in \mathbb{R}^{4}$, and the input vector $u \in \mathbb{R}^{6}$ are
\begin{eqnarray*}
	y&=&\left [ H_{2} \quad x_{A2} \quad x_{B2} \quad T_{2} \right ]^{T},\\
	u&=&\left [ H_{1} \quad x_{A1} \quad x_{B1} \quad T_{1} \quad F_{20} \quad Q_2 \right ]^{T},
\end{eqnarray*}
where $Q_{2}$ is the external heating, $F_{20}$ is the input flow of product A, and $H_1, x_{A1}, x_{B1}$ and $T_{1}$ are the states associated to the first reactor. 
The values and units of the parameters of the benchmark system are collected in Table \ref{table:system_parameter}.
For a more detailed description of the system the interested reader is addressed to \cite{stewart2010cooperative}.

\begin{table}
	\caption{Parameters of the chemical benchmark system}
	\label{table:system_parameter}
	\begin{center}
		\begin{tabular}{c c c|c c c}
			\toprule
			Parameter& Value& Unit &Parameter&Value&Unit\\
			\midrule
			$\rho$&0.15 &$kg/m^3$ &$A2$&3&$m$\\
			$kv_1, kv_2$&0.5&$kg/m \ s$&$x_{A_{0}}$&1&$wt\%$\\
			$k_A$&0.336&$1/s$&$k_B$&0.089&$1/s$\\
			$E_A/R$&-100&$kJ/kg$&$E_B/R$&-150&$kJ/kg$\\
			$\Delta H_A$&-40&$kJ/kg$&$\Delta H_B$&-50&$kJ/kg$\\
			$C_p$&2.5&$kJ/kg\ K$&$T_0$&313&$K$\\
			\bottomrule	
		\end{tabular}
	\end{center}
\end{table}

\subsection{Training of an LSTM model of the system}
\begin{figure}
	\centering
	\includegraphics[width=0.725\linewidth]{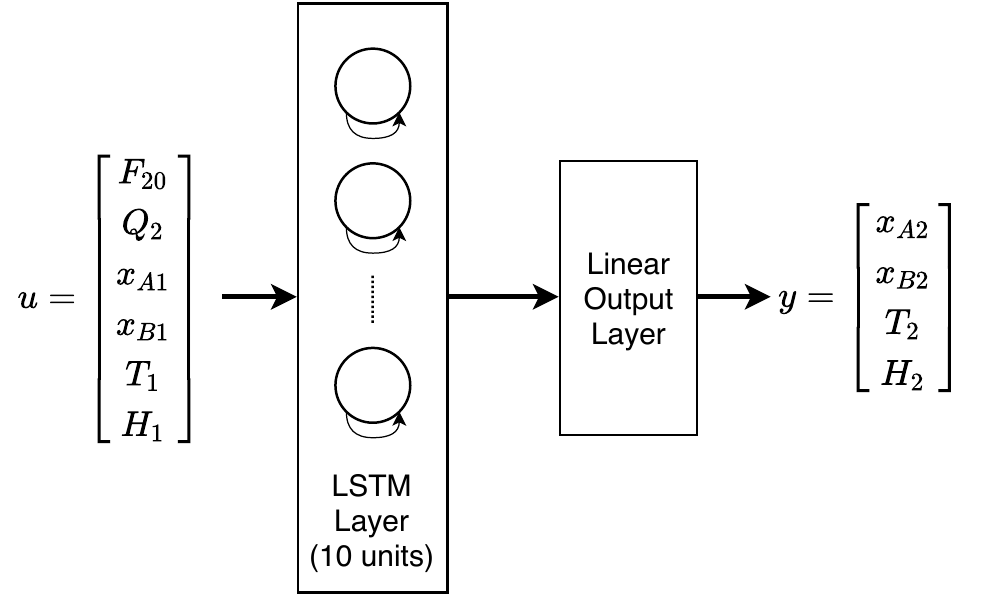}
	\centering
	\caption{Structure of one-layer LSTM network.}
	\label{fig:LSTMnet}
\end{figure}

\begin{figure}
	\centering
	\includegraphics[clip, trim=15mm 2mm 40mm 15mm, width=0.9\linewidth]{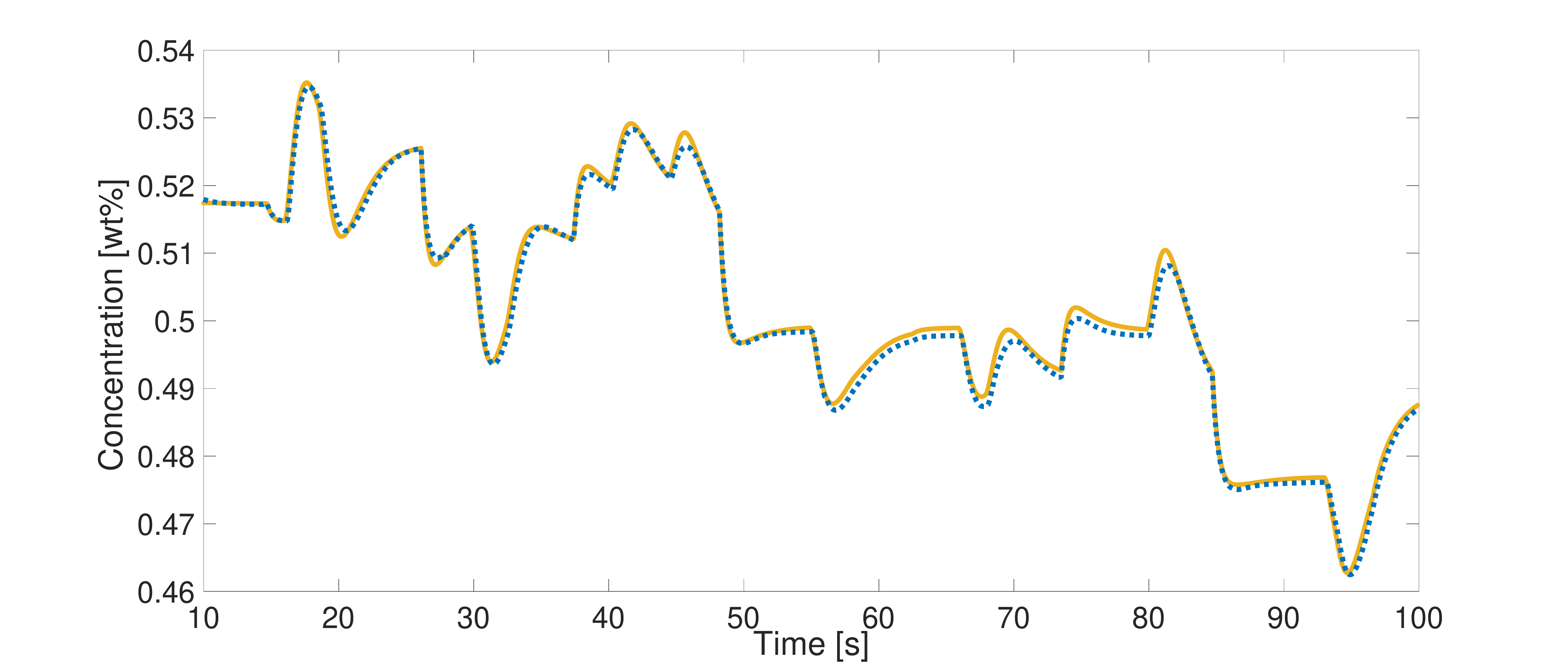}
	\centering
	\caption{Open-loop prediction of output $x_{A2}$  by the LSTM model (blue dotted line) compared to the ground truth (yellow solid line).}
	\label{fig:LSTMperformxa2}
	\vspace{2mm}
	\includegraphics[clip, trim=15mm 2mm 40mm 15mm, width=0.9\linewidth]{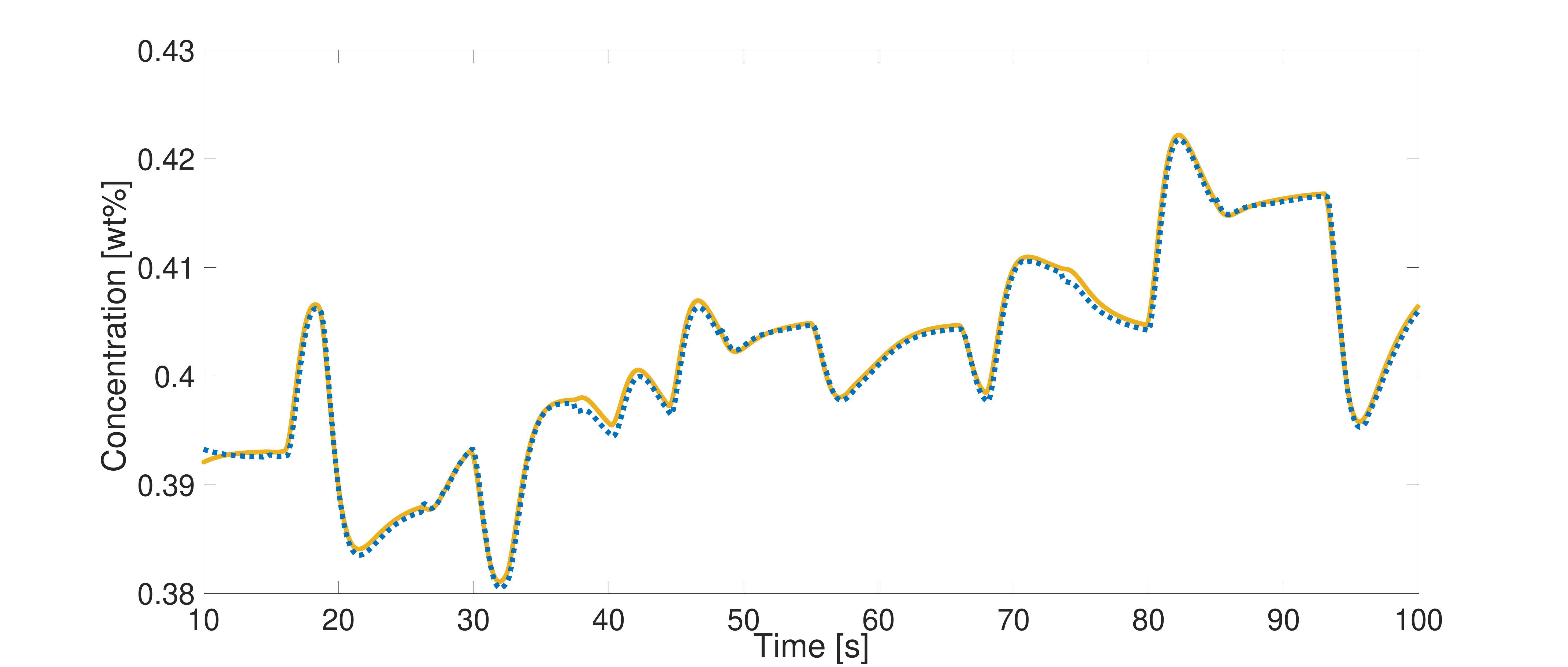}
	\centering
	\caption{Open-loop prediction of output $x_{B2}$  by the LSTM model (blue dotted line) compared to the ground truth (yellow solid line).}
	\label{fig:LSTMperformxb2}
\end{figure}

An LSTM network is used to learn the chemical system in nominal conditions.
To this end, \REV{following the guidelines described in \cite{terzi2021learning}} a single-layer LSTM network with $10$ neurons and a linear output transformation has been adopted, with the structure depicted in Figure \ref{fig:LSTMnet}. 
A total of $136$ input-output sequences has been collected from the simulated system \eqref{eq:benchmark}, with sampling time $\tau = 0.1s$ and length $T_{s} = 100s$.
Of these sequences, $100$ have been used to train the LSTM network by minimizing the Mean Squared Error (MSE) between the collected outputs and the open-loop network's prediction, as discussed in \cite{terzi2021learning}.
The remaining $36$ sequences have been used to independently test the LSTM model's performances.
In Figures \ref{fig:LSTMperformxa2} and \ref{fig:LSTMperformxb2} the LSTM model's open-loop prediction of the outputs $x_{A2}$ and $x_{B2}$ are compared to the real output trajectories, respectively, for one of the test sequences. 
Note that the first 10 seconds of prediction are discarded, as they constitute the so-called washout period \cite{terzi2021learning}.

\begin{figure}[t]
	\centering
	\includegraphics[clip, trim=15mm 0mm 40mm 15mm, width=0.9\linewidth]{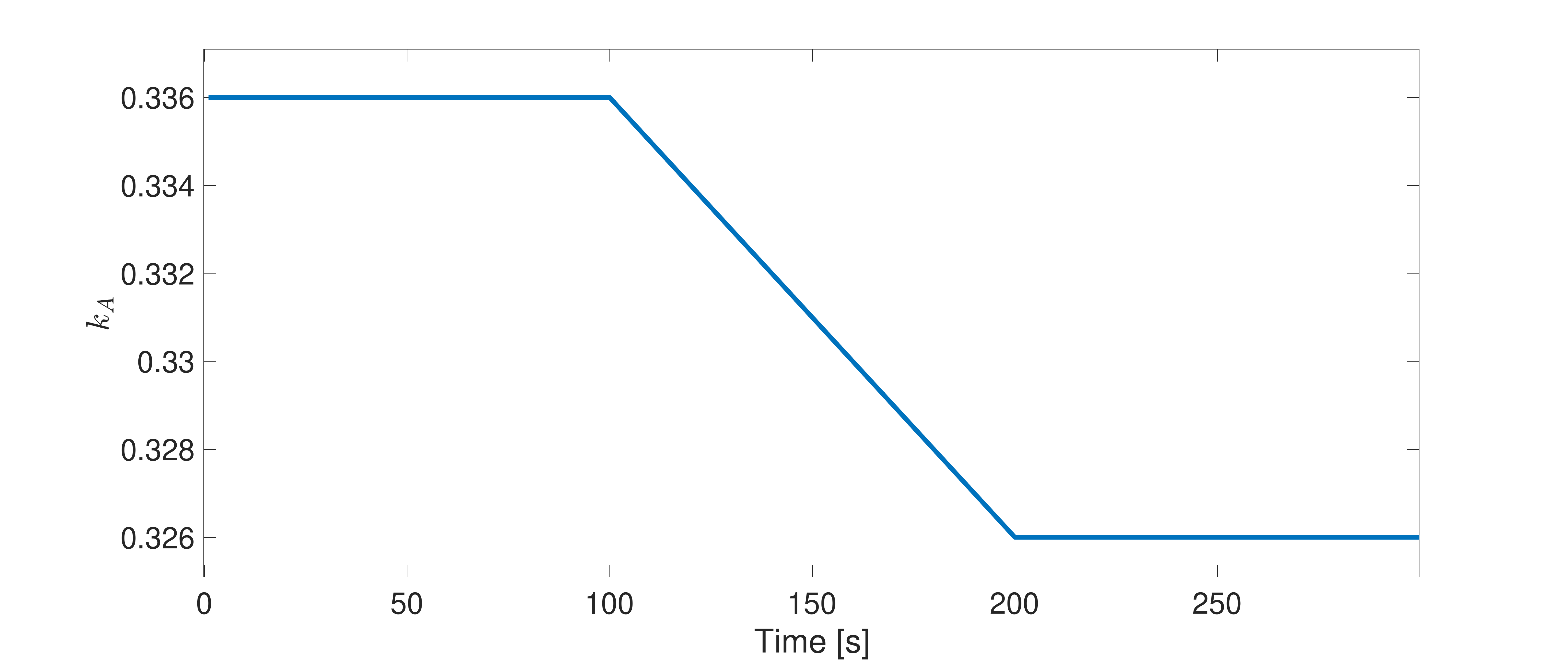}
	\centering
	\caption{Drift of parameter $k_A$.}
	\label{fig:kadrift}
\end{figure}

\subsection{Parameter drift} \label{sec:banchmark:drift}
In practice, a common phenomenon is that, during the operation of the plant, the value of some constant parameter varies for different reasons.
For example, we consider the case where the parameter $k_A$ drifts from value $k_A = 0.336$ to value $k_A=0.326$ in the time interval $t\in[100,200]$, as shown in Figure \ref{fig:kadrift}. \REV{Note that one may also consider the case where multiple parameters of the system change at the same time.}
%
The performance degradation of the LSTM model due to the occurrence of the drift is measured by collecting other $35$ input-output sequences from the plant, and computing the MSE between the drifted system's output and the open-loop prediction of the model.
The results, reported in Table \ref{table:LSTM_performance_original}, witness a significant performance degradation.
In particular, the MSE increases from $1.43 \cdot 10^{-5}$ to $31.48 \cdot 10^{-5}$,  due to a degradation of outputs $x_{A2}$ and $x_{B2}$ modeling performances.
\begin{table}
	\caption{Modeling performances of the unadapted LSTM}
	\label{table:LSTM_performance_original}
	\begin{center}
		\begin{tabular}{c|cccc|c}
			\toprule
			& \multicolumn{5}{c}{MSE [$\cdot 10^{-5}$]} \\
			Scenario& $H_{2}$ &$x_{A2}$&$x_{B2}$&$T_{2}$ &  Average \\
			\midrule
			before drift&0.78 &1.36&1.06&2.53&1.43\\
			after drift&0.77 &46.74&77.16&2.72& 31.84 \\
			\bottomrule	
		\end{tabular}
	\end{center}
\end{table}

\subsection{MHE for LSTM parameter tuning}
To mitigate for the impact of the parameter drift on the modeling performances, the LSTM model needs to be tuned on-line.
To this end, we implemented the algorithm proposed in Section \ref{sec:problem} \REV{where, since the network state is not measured, a state observer has been used \cite{terzi2021learning}}.
Hence, during the parameter drift the LSTM weights have been tuned by means of the MHE algorithm. 
The \REV{nonlinear} optimization problem \eqref{eq:mhe_formulation} consists of $724$ optimization variables, i.e. the network's weights, \REV{and it has been solved with  MATLAB's \emph{fmincon}}.

Different values of the MHE hyperparameters $\mu$ and $N$ have been tested, validating the modeling performances of the resulting (tuned) LSTM model on the same $35$ sequences on which, in Section \ref{sec:banchmark:drift}, the performance degradation of the nominal LSTM model has been assessed.

\begin{figure}
	\centering
	\includegraphics[clip, trim=15mm 2mm 40mm 15mm, width=0.9\linewidth]{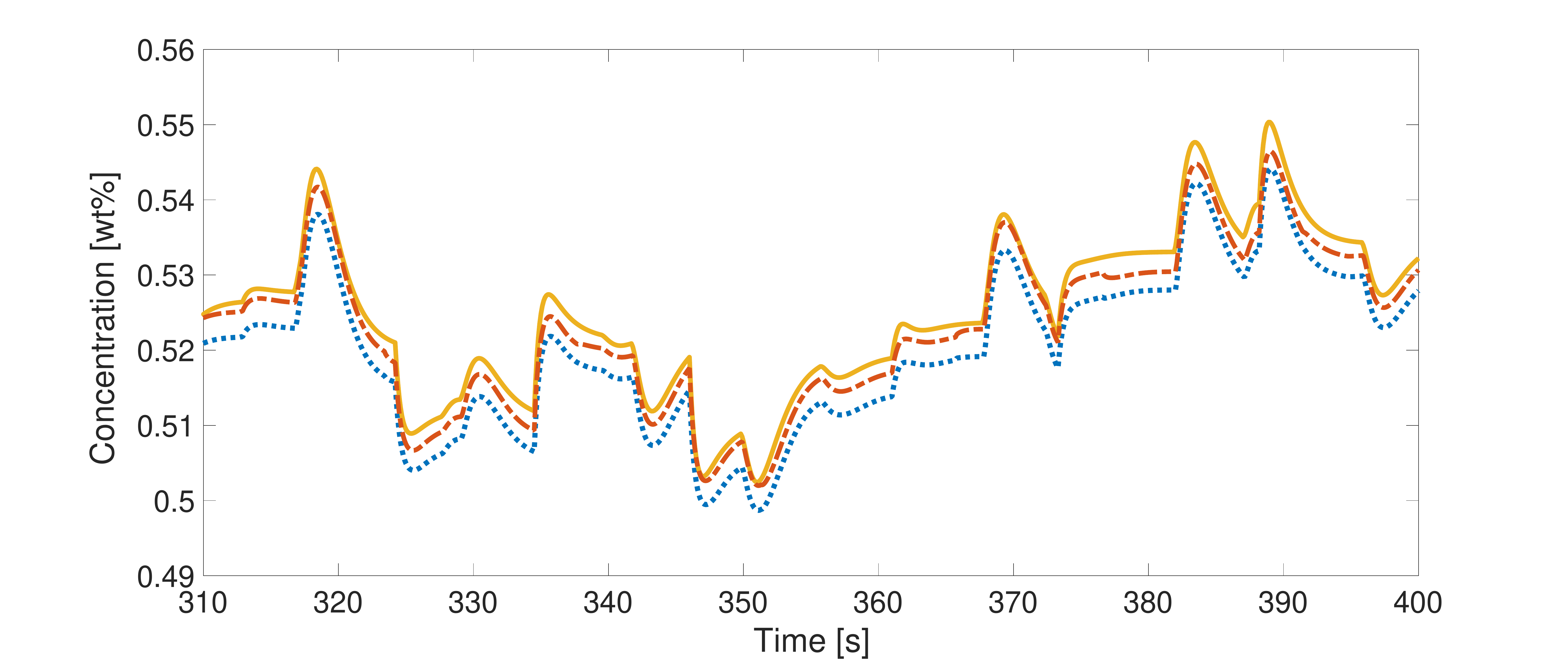}
	\centering
	\caption{Open-loop prediction of $x_{A2}$ by the unadapted LSTM model (blue dotted line) and tuned LSTM model (red dashed line) compared to the ground truth (yellow solid line), with $\mu=0.1$ and $N=10$.}
	\label{fig:comparisonxa2}
	\vspace{2mm}
	\includegraphics[clip, trim=15mm 2mm 40mm 15mm, width=0.9\linewidth]{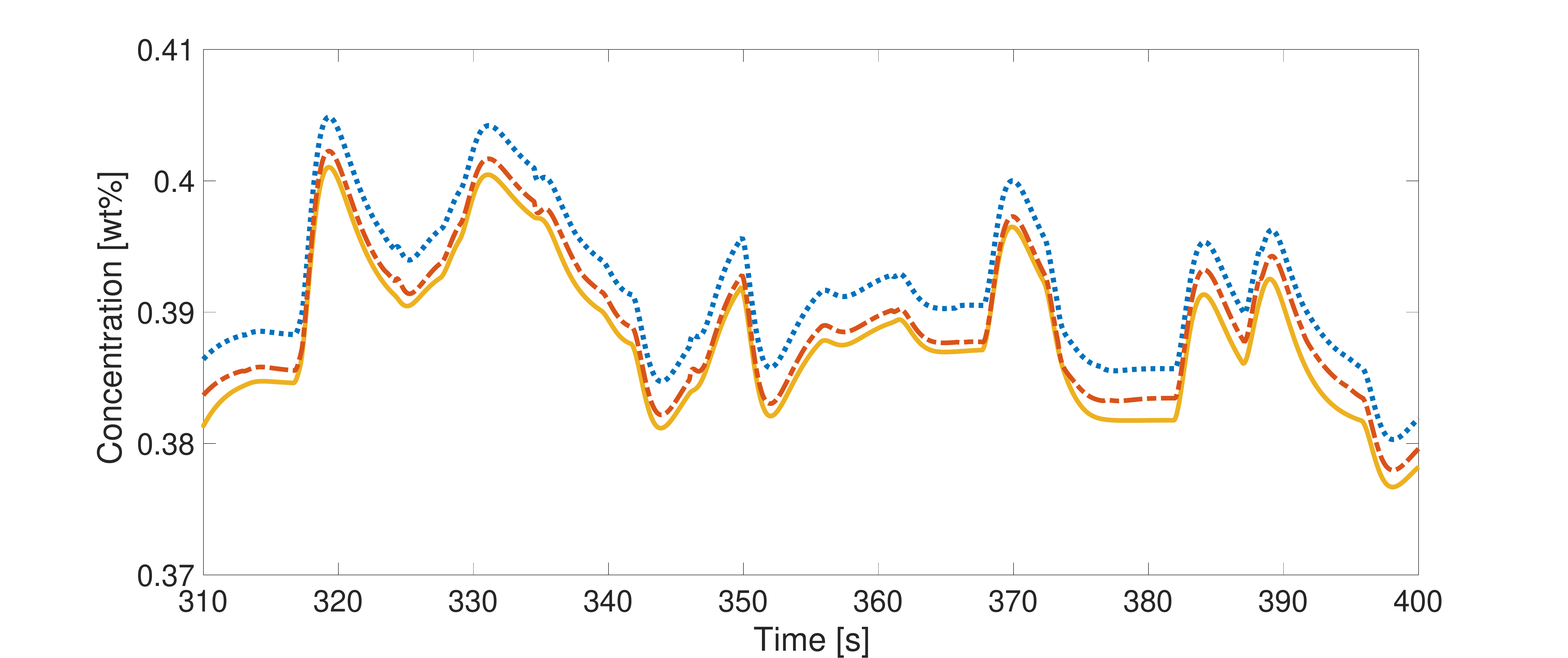}
	\centering
	\caption{Open-loop prediction of $x_{B2}$ by the unadapted LSTM model (blue dotted line) and tuned LSTM model (red dashed line) compared to the ground truth (yellow solid line), with $\mu=0.1$ and $N=10$.}
	\label{fig:comparsionxb2}
\end{figure}

In Table \ref{table:LSTM_performance_tuned} \REV{the computational burden and} the simulation error achieved by the tuned LSTM models are reported, witnessing a remarkable reduction of the MSE compared to that scored by the nominal (unadapted) LSTM model (see Table \ref{table:LSTM_performance_original}).
In particular, the best performances are achieved selecting $\mu=0.1$ and $N=10$, that allow to reduce the MSE from $31.84 \cdot 10^{-5}$ to $7.24 \cdot 10^{-5}$, i.e. a $77\%$ reduction.
The corresponding open-loop prediction of $x_{A2}$ and $x_{B2}$ are reported in Figure \ref{fig:comparisonxa2} and  Figure  \ref{fig:comparsionxb2}.

\begin{table}
	\caption{MHE parameters and tuned LSTM's modeling performances}
	\label{table:LSTM_performance_tuned}
	\begin{center}
		\resizebox{\columnwidth}{!}{
		\begin{tabular}{cc|c|cccc|c}
			\toprule
			\multicolumn{2}{c|}{Parameters} &  &
			\multicolumn{5}{c}{MSE [$\cdot 10^{-5}$]} \\
			$\mu$ & $N$ & Comp. time $[s]$ & $H_{2}$ &$x_{A2}$&$x_{B2}$&$T_{2}$ & Average \\
			\midrule
			0.05 & 10 & 33.6 & 2.19 & 9.01 & 14.28 & 3.99 & 7.37\\
			0.1 & 5 & 13.7 & 2.32 & 8.62 & 15.33 & 6.89 & 8.29\\
			0.1 & 10 & 24.8 & 1.81 & 9.33 & 13.6 & 4.26 & \textbf{7.24}\\
			0.1 & 20 & 39.6 & 3.23 & 8.09 & 19.55 & 8.00 & 9.72\\
			0.5 & 10 & 17.45 & 1.99 & 8.79 & 16.28 & 6.69 & 8.44\\
			\bottomrule	
		\end{tabular}}
	\end{center}
\end{table}
\section{Conclusions and future directions}\label{sec:conclusion}
In this paper, a possible solution to the lifelong learning problem for RNNs used as dynamic systems models is presented.
The theoretical convergence  properties of the proposed method, which is based on a Moving Horizon Estimation strategy, have been analyzed, and its performances have been evaluated on a nontrivial simulated chemical plant, already widely used as benchmark in the control community.

\REV{Notably, although it is assumed that no measurement noise affects the system, white gaussian noise could be easily included while preserving the boundedness of the weight estimation error. }

While promising results have been achieved, there are many open questions that need to be investigated and that may draw future research directions.
First and foremost, the state is herein assumed to be known. While this is true when NNARX are used, this is not the case for other RNN architectures, which instead call for the design of a suitable state observer.
Albeit, from a theoretical point of view, one could simply extend the state equations with fictitious dynamics of the parameters \cite{alessandri2008moving, kuhl2011real, robertson1996moving}, this approach could lead to an intractable optimization problem.
A possible solution towards this issue would exist if the state estimation is unconstrained, and it would consist of a combination of the proposed method with a Luenberger observer.

A second problem concerns the scalability of the optimization problem here proposed.
Indeed, Neural Networks are generally characterized by a possibly large number of weights.
For non-small NNs, this may lead to a large computational burden, or even intractability, although relieving strategies, such as batch operation, could be adopted.


\section*{Acknowledgments}
\hspace{0.8cm}
 
\begin{minipage}[c]{0.125\textwidth}
	\includegraphics[width=\textwidth]{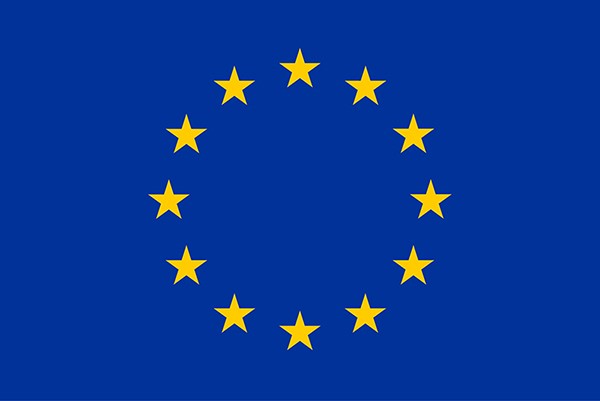}
	\label{fig:euflag}
\end{minipage}
\hspace{0.2cm}
\begin{minipage}[right]{0.325\textwidth}
	This project has received funding from the European Union’s Horizon 2020 research and innovation programme under the Marie Skłodowska-Curie grant agreement No. 953348
\end{minipage}
\\
\bibliographystyle{IEEEtran}
\bibliography{Bibliografia}

\end{document}